\documentclass[graybox]{svmult}

\usepackage{type1cm}        
\usepackage{makeidx}         
\usepackage{graphicx}        
\usepackage{multicol}        
\usepackage[bottom]{footmisc}

\usepackage{newtxtext}       %
\usepackage[varvw]{newtxmath}       

\makeindex             

\usepackage{natbib}
\usepackage{ulem}

\def\hho{H$_2$O}
\def\coo{CO$_2$}
\def\soo{SO$_2$}
\def\meth{CH$_4$}
\def\oo{O$_2$}
\def\ooo{O$_3$}
\def\nn{N$_2$}
\def\hh{H$_2$}

\def\degc{$^0$C}

\def\gcc{g\,cm$^{-3}$}

\begin{document}

\title*{Likelihood and appearance of life beyond the Earth: An astronomical perspective}
\titlerunning{Life beyond the Earth: An astronomical perspective} 
\author{Floris van der Tak\orcidID{0000-0002-8942-1594}}
\institute{Floris van der Tak \at Space Research Organization Netherlands (SRON) \& Kapteyn Astronomical Institute, University of Groningen, Landleven 12, 9747 AD Groningen, The Netherlands; \email{vdtak@sron.nl}}
%
\subtitle{\it Book chapter, to appear in First Contact: Aliens and Humans in Contemporary Science Fiction, eds. M. van Dijk, F. Bosman, K. Glimmerveen (Springer)}
%
\maketitle


\abstract{
As of 2025, over 6000 planets are known to orbit stars other than our Sun. 
We can measure their sizes and orbital periods, infer their masses and temperatures, and constrain their compositions. 
Based on these data, about 1\% of extrasolar planets are potentially habitable for life as we know it, implying that of the billions of planets in our Galaxy, some may actually be inhabited, at least by microbes. 
However, recognizing signs of alien life forms is a major challenge for current technology, because of the wide range of conditions on extrasolar planets, and because of the wide range of forms that life may take.
This chapter reviews observations of exoplanets and discusses astrobiological definitions of habitability and the likelihood of finding life beyond the Earth, both within and outside the Solar system.}

\bigskip


\section{Introduction}
\label{s:intro}


The world that we live on is a small rocky planet, which orbits a mid-size star. 
Our solar system contains four small rocky planets orbiting the Sun, and four gas giants orbiting further out. 
The planets vary greatly in their sizes, masses, and compositions, and no two planets are alike.
The planets orbit the Sun in almost the same plane, in the same direction, suggesting that the Solar system formed as a whole, in a rotating flat disk called the Solar nebula (\citealt{swedenborg1734}, \citealt{kant1755}).
In addition to the Sun and the major planets, the Solar system contains at least nine dwarf planets\footnote{Like a planet, a dwarf planet orbits a star and is massive enough to assume a nearly round shape, but unlike a planet, it is not massive enough to clear its neighbourhood.}, over 100 moons, thousands of known (and many more unknown) comets, millions of asteroids, and dust grains, visible as the zodiacal light just before sunrise or just after sunset.
The ordering of these bodies by mass is opposite of that by number: the Sun contains 99\% of the mass of the Solar system, the planets 1\%, while the other bodies contribute a negligible amount of mass.

The idea that other stars may also have orbiting planets goes back at least to \citet{bruno1584} and \citet{newton1713}, and many searches were undertaken in the nineteenth and twentieth centuries.
Early searches for extrasolar planets focused on astrometric techniques, where the orbiting planet causes a periodic motion of its host star in the plane of the sky.
Such motions are visible with digital cameras on space-based telescopes \citep{gaia2023}, but they are too small to be seen from the ground with photographic plates.
Unfortunately, several false claims of exoplanet detections (which turned out to be measurement errors or low-mass stellar companions) appeared until the 1970s, giving the field a bad reputation. 
The first planet around a solar-type star was found by the periodic radial motion of its host star seen through the Doppler effect\footnote{The Doppler effect shortens the wavelength of sound and light from objects approaching the observer, leading to blueshift for light and higher pitch for sound. The opposite effect occurs for receding objects.}, a technique borrowed from the searches for low-mass binary stars \citep{mayor1995}. 
Their discovery led to the Nobel Prize in Physics 2019.
The radial motion is large if the planet is massive and its orbit is small, so naturally, the first planets found this way have masses like Jupiter but orbits like Mercury.
Nicknamed hot Jupiters, these planets form a population without analog in our Solar system.

The radial velocity observations led to predictions that some planets would pass in front of their host stars in a so-called transit, something occurring for Mercury every $\sim$10 years and for Venus every $\sim$100 years. 
The first transiting exoplanet was found by \citet{charbonneau2000}, and ESA's Corot mission (2007--2014) and especially NASA's Kepler mission (2009--2018) have turned this method into an industry yielding thousands of exoplanets. 
Unlike the radial velocity method, the transit method gives unambiguous sizes of the planet and its orbit. 
Like the radial velocity method, transits are biased toward large planets on small orbits. 

The first direct detections of exoplanets were made in 2008 with the Keck and Gemini telescopes \citep{marois2008}.
This technique is currently only feasible for large planets which are young and warm enough to be bright in the infrared.
The main challenge is achieving a high contrast in the image, to avoid the star outshining the planet.
Space observatories have an advantage here: NASA's Roman observatory, to be launched 2027, will be able to detect mature gas giants.
Other techniques to find exoplanets include the use of gravitational lensing; see \citet{perryman2018} for a review.

As of October 2025, these techniques have together revealed over 6000 exoplanets \citep{xpl-arxiv}. 
The question if some of the planets within or beyond our Solar system harbour life goes back centuries, with recorded suggestions going back to e.g. \citet{huygens1698} and \citet{herschel1795}. 
This paper reviews the status of the search for habitable worlds in the Universe from an astronomical point of view.
Section~\ref{s:xpl} describes the observed properties of the known exoplanet population.
Section~\ref{s:habit} reviews the concept of planetary habitability, and estimates the number of potentially habitable worlds.
Section~\ref{s:bottles} discusses limits to planetary habitability.
Section~\ref{s:search} describes searches for life beyond the Earth.
Section~\ref{s:concl} draws conclusions and outlines future prospects in this field.

\section{The observed exoplanet population}
\label{s:xpl}

After 30 years of exoplanet research, it is clear that planets are more common than stars.
On average, there are at least 1.1 planets per star, but the distribution depends on mass and composition.
While only one in 10 stars has a Jupiter-size planet, 1 in 3 has a Neptune-sized planet, and every star (on average) has an Earth-sized planet.
Stars with higher fractions of elements heavier than helium tend to have more gas giant planets, which is consistent with scenarios where planetary cores are built up from heavy elements, although the trend does not hold up for terrestrial planets \citep{buchhave2012}.

Minor bodies around other stars are also known: extrasolar comets were discovered around the star $\beta$~Pic in 1987, before the first exoplanet \citep{ferlet1987}, and have been found since around $\sim$30 stars by spectroscopy and/or photometry\footnote{Photometry is measuring the absolute intensity of a light source; spectroscopy is measuring the distribution of the light over wavelengths.} \citep{strom2020} and recently also by their transits \citep{zwieba2019}.
Belts of asteroids and dust grains are seen in near- and far-infrared images of nearby stars \citep{greaves1998,kalas2005}, and ring systems may have been seen around two exoplanets \citep{kenworthy2015,kenworthy2023}.
So far only two moons may have been found around exoplanets (e.g., \citealt{kipping2022}), but both claims are disputed.
This low number is surprising since moons outnumber planets 10:1 in the Solar system, and the largest local moons are bigger than the smallest planets. 
Direct detections of exoplanets will help, e.g. with the Roman telescope, since they allow to observe exomoons in transit across their host planets. 
Dynamical stability arguments suggest that exomoons may mostly occur around long-period planets, which require observational campaigns of a decade or more to discover \citep{dobos2022}.
Such campaigns are also needed to find true analogs of our own Solar system.
 
\begin{figure}[b]
\centering
\hspace*{-1cm}
\includegraphics[width=1.19\textwidth]{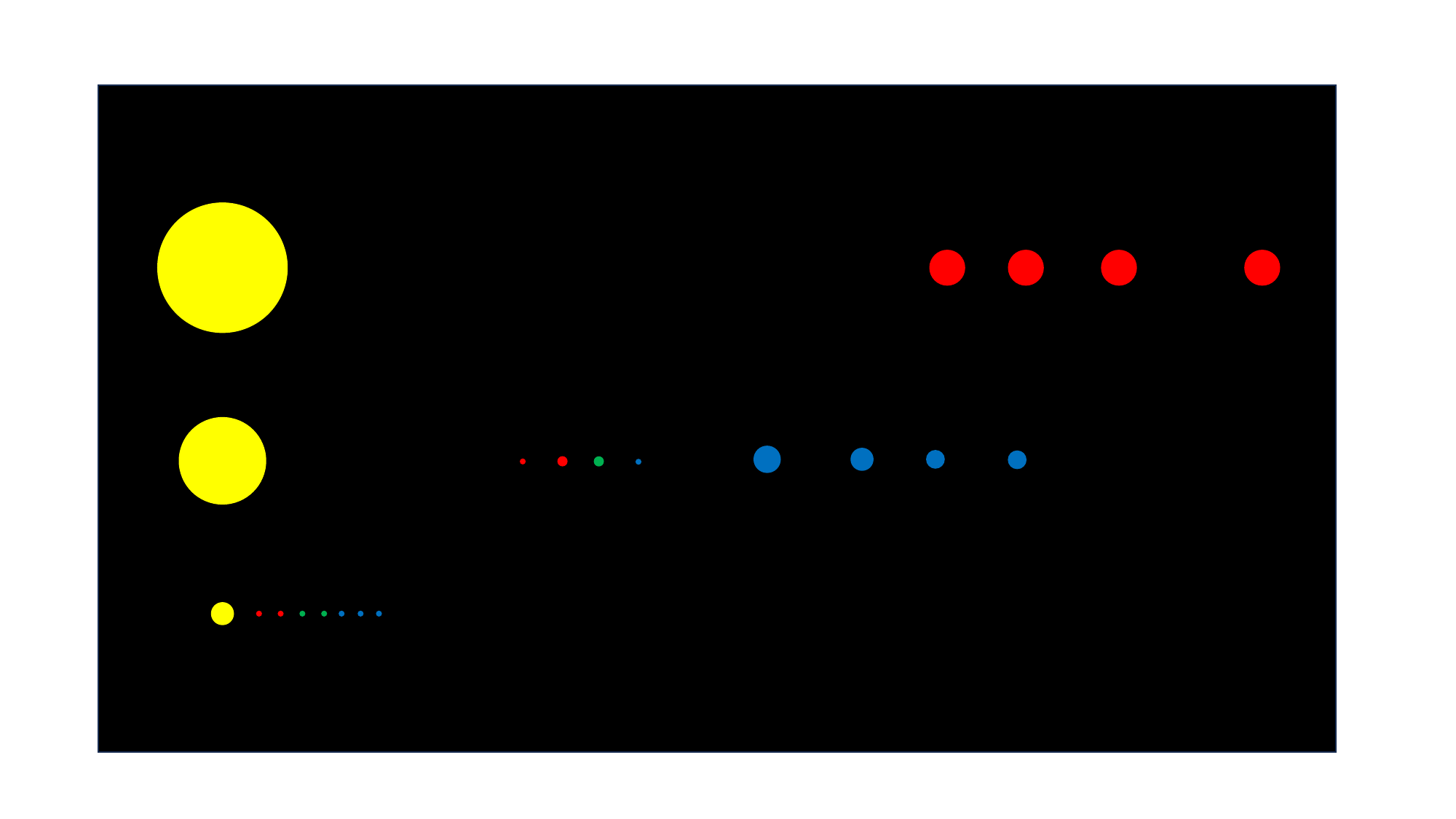}
%
%
\caption{Relative sizes of three planetary systems: HR 8799 (top), the Solar system (middle) and TRAPPIST-1 (bottom). Dot sizes correspond to relative masses, and horizontal distances to orbit sizes, both on logarithmic scales. Stars are yellow, hot planets are red, planets inside the habitable zone are green, and cold planets are blue. For the habitable zone limits, the `loose' definition was adopted; see Section~\ref{s:habit}.}
\label{fig:planets}
\end{figure}

The masses and sizes of exoplanets are among their most fundamental properties.
The radial velocity technique only gives a lower limit to the planet mass, because the inclination of the orbital plane to our line of sight is unknown.
If the planet is also transiting, the orbit must be edge-on, so the mass is known, and the transit depth gives the size.
Modern techniques allow us to measure radial velocities below 1 m/s (corresponding to a Doppler shift of 10$^{-9}$ of the wavelength) and transit depths of $10^{-5}$ in relative brightness. 
The masses of known exoplanets range from several Jupiter masses to below an Earth mass; see 
\citet{xpl-arxiv}
for more information.
Figure~\ref{fig:planets} compares the architecture of the Solar system to that of two well-known exoplanet systems.

The searches for exoplanets have revealed several planet types that do not occur in the Solar system.
In particular, most planets in the Universe turn out to be intermediate size, i.e. with a mass between that of Neptune and the Earth.
Planets orbiting very close to their host stars are another example of planets without local analog. 
Gas giants are unlikely to form close to their host stars, so the discovery of hot Jupiters indicates that planets may migrate to other orbits after their formation. 
This type of planet, while easily visible, is actually quite rare, with an occurrence rate below 1\%.
On the other hand, direct imaging has revealed planets very far from their host stars (several times as far as in the Solar system), and even planets without host stars, such as the Jupiter-sized free-floating planets in Upper Sco \citep{miretroig2022}.
Orphan planets may result from dynamical ejection from multiple systems, or from formation processes like those of low-mass stars.

The temperatures of exoplanets are essential input to know their physical state.
Planetary temperatures can be roughly estimated by assuming an equilibrium between heating by starlight and cooling by thermal radiation.
The temperature then mainly depends on the stellar luminosity and the orbital size, which are measured, and on planetary albedo and emissivity, for which fiducial values must be adopted.
This method depends on the assumed rotation rate, and ignores the greenhouse effect (noticeable for Earth, strong on Venus) and heat from the planetary interior, both as leftover from the formation (important for Jupiter and Saturn), and from the decay of radioactive elements.
The necessary input data to estimate these effects are usually not available for exoplanets, but reasonable assumptions can be made to construct models. 
Temperatures derived this way range from below the boiling point of nitrogen to above the melting points of rocks and metals.
The warmest Earth-sized planets and the majority of detected intermediate-size planets are therefore thought to possess magma oceans on their surfaces, similar to the Earth in its early days.
In the future, imaging of exoplanets at multiple infrared wavelengths will allow us to directly measure their emission temperatures.

The bulk densities of exoplanets are important as clues to their composition.
For transiting planets, the mass is known from the radial velocity amplitude, and the size is known from the transit depth. 
Combining mass and size yields the bulk density, which is found to range from 0.2 \gcc, corresponding to pure hydrogen/helium gas (at room temperature and pressure), to 10 \gcc, similar to pure solid iron (ignoring compression effects).
Intermediate densities would correspond to pure water (1 \gcc) or rock (3--5 \gcc), but such planets probably have a mixed composition, like the Earth with its metallic core and rocky mantle.
Further clues to the planetary composition may be obtained from spectroscopy of its atmosphere (for volatile species) and its host star (for refractory species).

Studying planetary atmospheres is useful to learn about planetary (volatile) composition, as atmospheres are more accessible than surfaces or interiors, although atmospheric composition is not necessarily representative of planetary composition, as the example of the Earth shows.
Studying atmospheres is also useful to learn about planetary formation (e.g., from dust or from gas) and subsequent evolution (e.g., migration to other orbits, or pollution by collisions with other bodies).
Third, studying atmospheres gives clues to planetary habitability, since an atmosphere allows to retain volatile material (important for the circulation and transport of organics), by its condensation into clouds.
In our Solar system, all planets and several moons possess atmospheres, but their properties vary widely.
Jupiter and Saturn have thick atmospheres of Solar composition (mostly hydrogen and helium), indicating that they formed with the Solar system.
The atmospheres of Uranus and Neptune, while also thick and hydrogen-helium containing, consist mostly of condensable gases (NH$_3$, \hho, \meth), indicating that their formation ended when the Solar system lost its gas.
In contrast, the rocky planets have thin heavy atmospheres, consisting of \coo\ (Venus and Mars) or \nn\ (Earth; and also Saturn's largest moon Titan).
These planets lost their initial hydrogen-helium atmospheres (as evidenced by the depletion of heavy noble gases) and gained their current 'secondary' atmospheres by outgassing from magma oceans \citep{kite2020} and delivery by comets and asteroids \citep{broadley2022}.
After their formation, they evolved significantly: Mars and Venus lost their water, and the Earth atmosphere was enriched in \oo\ by biogenic processes.
For a review of volatile segregation and processes during planet formation, see \citet{suer2023}.

The atmospheres of exoplanets can be studied if the planet is seen in transit or in direct detection.
For transiting planets, the atmospheric composition can be obtained by comparing the spectrum during transit (when the starlight is filtered by the planetary atmosphere) with that during eclipse, when the planet passes behind the star so that pure starlight is observed.
Between transit and eclipse, a mix of direct and reflected starlight is observed in the visible part of the spectrum, and a mix of stellar and planetary emission in the infrared. 
By tracking this mix along the orbit (in a so-called phase curve), longitudinal structure can be observed, such as differences between the morning and evening sides of the planet.
Such information can also be obtained from time-resolved transit observations, in particular during ingress and egress.
For an overview of these techniques see \citet{crossfield2015} and \citet{deming2019}; for an overview of results see \citet{madhusudhan2019}.
About a dozen atomic and a dozen molecular species have been found prior to the launch of the James Webb Space Telescope (JWST) in December 2021.
Highlights of JWST results in this area include the detection of \soo\ in WASP-39b (a sign of active photochemistry; \citealt{tsai2023}), and the simultaneous detection of \coo\ and \meth\ in K2-18b \citep{bell2023}.
So far such measurements are only possible for gas giants, although for lower-mass planets, first results are appearing \citep{hu2024,ducrot2025}.
Quantitative analysis of exoplanet atmospheres remains difficult: estimated abundances depend strongly on the model for the background atmosphere.

\section{Definitions of planetary habitability}
\label{s:habit}

To specify the conditions that a planet needs to harbour life, we look into the basic needs of living organisms, which requires an operational definition of what life is.
This review focuses of life as we know it to exist on Earth, since searching for something unknown is harder than searching for something familiar.
Searches for alternative forms of life are being undertaken; for a recent account see e.g. \citet{cleaves2023}.

While plants and animals form the most visible kinds to life to us, the most common form of life on Earth are bacteria and other microbes. 
The basic unit of life as we know it is the cell, which performs metabolism and reproduction, and can undergo (Darwinian) evolution.
To perform these functions, life as we know it has four basic requirements. 
The first is energy, which may come in the form of heat (e.g. on the ocean floor; essentially a leftover from planetary formation), sunlight (as used by plants and other photosynthetic organisms), chemical energy (as used by some bacteria), or existing organic matter (as used by animals).

The second requirement are nutrients, for example to build cell membranes (for compartmentalization) and information carriers (for reproduction). 
Biomolecules on Earth primarily rely on six chemical elements: carbon, hydrogen, nitrogen, oxygen, phosphorus, and sulphur (collectively abbreviated as CHNOPS). 
For example, the DNA molecule contains all these elements except sulphur.
Many organisms depend additionally on other elements, such as iron to bind \oo\ in blood, or molybdenum to liberate nitrogen from \nn, but these are not universal needs.
Life's need for transition metals as catalysts highlights the importance of planetary processes such as limited differentation and late accretion \citep{wade2021}.

The CHNOPS set of elements is far from the set of most common elements in the Solar system: the latter is dominated by the Sun which consists for 99\% of hydrogen and helium \citep{lodders2003}.
The set does also not consist of the most common elements on Earth, which would include iron and silicon \citep{mcdonough2008}, nor those dominating the Earth crust \citep{rudnick2003}.
They are the most common volatile elements on Earth, which means that they occur primarily in compounds with low boiling and melting points.
They are the elements dominating the Earth's oceans, and they share a high electronegativity, so that they form soluble salts or molecules \citep{millero2008}.

Of the CHNOPS elements, only hydrogen has existed since the first three minutes of the Universe, when atoms formed out of subatomic particles.
Elements heavier than lithium were formed later by nuclear fusion at the centers of stars; elements heavier than iron were formed in supernovae, neutron star mergers, and other energetic events leading to high neutron fluxes.
In the so-called s- and r-processes, the capture of single resp. multiple neutrons is followed by $\beta$-decay into new elements.
Building up sufficient CHNOPS to form rocky planets with water and biomolecules has taken $\sim$2 billion years \citep{dayal2015}, which leaves $>$10 billion years for the development of life on planets.

Proposals for alternative biochemistries based on silicon instead of carbon occur both in the scientific and in the popular literature (such as the rock-eating Horta from Star Trek).
The idea is that, since the two elements are just below each other in the periodic table, their chemistries should be similar.
From an abundance point of view this is possible: while in the Sun, silicon is 16 times rarer than carbon, the difference is only a factor of 2 on Earth.
The problem with silicon-based life is the low volatility of its compounds, especially under oxidizing conditions. 
In particular, carbon oxides are gases, while silicon oxides are solid.  
Reducing (H-rich) conditions may be more favourable for silicon-based life; in particular, silane (SiH$_4$) is gaseous at room temperature and pressure.
In addition, silicon-silicon chains are less stable than carbon-carbon chains.
For further discussion see e.g. \citet{petkowski2020}.

The third requirement for life as we know it is a solvent. 
Only in solution, chemistry proceeds fast enough to build up complex molecules such as proteins. 
The liquid acts as a transporter which brings organics together. 
For effective mixing, water flows are important, which call for temperature gradients and/or tidal forces from a satellite.
The fourth and final basic requirement is the existence of stable benign conditions, which implies a stable orbit around a long-lived star without strong flares or other variability. 
The lifetimes of stars decrease rapidly with increasing mass, which renders stars significantly more massive than the Sun as inhospitable, as their lifetimes are less than a billion years.
The most common type of stars in our Galaxy are red dwarfs with masses of $\sim$0.5 solar masses and lifetimes of $>$10 billion years.
Many planets have been found around such stars, including our nearest neighbour Proxima b \citep{anglada2016} and the well-known TRAPPIST-1 system with seven temperate Earth-sized planets (\citealt{gillon2017}; Fig.~\ref{fig:planets} bottom).
However, the high flaring activity of red dwarfs may be a limit to the development of life on the planets orbiting them.

Other planetary conditions are probably helpful for life but may not be essential. 
These include the presence of a magnetic field, which shields against the Solar wind, cosmic rays, and other energetic particles from space which may cause atmospheric erosion. 
Another possibly important condition is the presence of plate tectonics and volcanism, which helps to stabilize the climate by cycling gases in and out of the atmosphere, in particular the greenhouse gas \coo\ \citep{oosterloo2021}.
Exactly which conditions are essential or helpful for the habitability of planets is a topic of ongoing discussions in the scientific literature \citep{cockell2016,meadows2018}.

Life's need for a solvent implies limits on the temperature and pressure at the planetary surface. 
In the case of water, a temperature between 0 and 100 \degc\ and a pressure above 6 mbar are needed to find \hho\ in its liquid form. 
Via the surface gravity, the pressure requirement translates into a size requirement for the planet to be at least Mars sized, although subsurface oceans can occur on smaller bodies (such as Europa).
By assuming radiative equilibrium, the temperature requirement translates into an allowed range of orbital sizes, which depends on the luminosity of the host star. 
This range is commonly known as the habitable zone (HZ) of a star \citep{kasting1993}, which holds for Earth-like atmospheric compositions.

While the original (sometimes called naive) concept of the HZ includes all orbits where the equilibrium temperature is between the freezing and boiling points of water, the range of orbit radii where liquid water may exist on the planet surface is considerably narrower, because of feedback effects \citep{kopparapu2013}.
In particular, the inner edge is limited by the 'runaway greenhouse' effect: evaporating water acts as a greenhouse gas, which leads to further heating and evaporation. 
This self-enhancing cycle has probably led to the hot and dry climate on Venus.
Similarly, the outer edge of the HZ is limited by the 'runaway snowball' effect. Ice caps reflect starlight, which leads to less heating and more freezing of water. 
This effect may have added to the cold dry climate on Mars, which had a water-rich surface until 2 billion years ago. 
On Earth, historic periods of glaciation were ended by increased heating due to periodic oscillations in its orbit known as the Milankovich cycles, as well as other effects such as overall solar brightening and volcanic \coo\ buildup.
Possibly the Earth was never fully frozen over but oceans remained open near the equator, a state known as the Jormungand climate \citep{abbot2012}.

The number of (potentially) habitable exoplanets may be estimated from the properties of known exoplanets, by applying limits to their sizes (for atmospheric pressure) and equilibrium temperatures, so that liquid water may exist on their surfaces. 
An upper limit to the number is obtained by taking the freezing and boiling points of pure water (273 and 373 K) as temperature limits; we call these the 'naive' limits.
More realistic limits including climate and feedback effects would be temperatures between 246--295 K (the 'loose' limits), or 260--289 K (the 'strict' limits) \citep{kopparapu2013}.
As of August 2023, 5502 exoplanet detections have been confirmed 
\citep{xpl-arxiv},
of which 378 have sizes between 0.4 and 2.5 Earth radii and temperatures within the 'naive' limits, of which 147 fall within the 'loose' limits, of which 98 within the 'strict' limits.
These numbers are approximate because the limits do not consider gases other than \hho, cloud effects, or 3D climate effects.

Depending on which limits are used, between 2\% and 7\% of planets are thus potentially habitable. 
Our galaxy, the Milky Way, contains $\sim$100 billion stars, and at least as many planets. Therefore, based on size and temperature, about a billion planets in the Milky Way are potentially habitable. 
Avoiding the Galactic Center, where the high rate of supernova explosions may limit habitability, and the Galactic outskirts, where the lower amount of heavy elements may impede planet formation, decreases this estimate by a factor of a few \citep{lineweaver2004}.
Only counting stars with lifetimes longer than the Sun makes an even smaller difference: the distribution of stellar masses drops off steeply toward the high-mass end.
We conclude that in our Galaxy alone, hundreds of millions of planets are potentially habitable, and the same holds in principle for most of the $\sim$100 billion other galaxies in the Universe. 
In addition, some exomoons may have the sizes and temperatures to allow liquid water on their surfaces.

Alternative biochemistries may use liquids other than water as a solvent. 
In particular, ammonia and methane may play this role, which are made of elements as abundant as those forming water.
The Cassini mission has found lakes of liquid methane and ethane on Saturn's moon Titan, with depths of over 100 m \citep{mastrogiuseppe2019}.
The advantage of water as a solvent over ammonia and methane is that it is liquid at a higher temperature (which speeds up chemistry) and over a larger temperature range (which widens the HZ).
The polar nature of water is a further advantage for life, as it helps to give structure to biomolecules, for instance cellular compartments, and the 3D structure of proteins.
However, on planets and moons lacking water, life may use alternative solvents, possibly at a lower efficiency; see \citet{bains2021} for further discussion.
Including this possibility would increase the number of habitable planets considerably.
Assuming that life will emerge with a high probability if conditions are suitable for its survival, we conclude that extraterrestrial life should be abundant, based on basic physical and chemical considerations.

\section{Bottlenecks for life on other planets}
\label{s:bottles}

Our estimate above that many millions of planets in the Galaxy are habitable does not mean that many planets are inhabited by plants and animals.
Even if they are inhabited, the most likely type of life are microbes, the simplest forms of life, which dominate the biomass on Earth. 
Many places on Earth (such as deserts and hot springs) only support certain microbial life forms (so-called extremophiles), and even our own bodies contain more bacteria (by number) than human cells \citep{merino2019}.
The dominance of microbial life was even stronger in the past: the oldest traces of life on Earth are traces of prokaryotes as old as 3.4--3.9 billion years \citep{javaux2019}, which is 90\% of Earth history. 
In contrast, eukaryotic cells developed 2.1 billion years ago (50\%), animals 700 million years (15\%); and Homo sapiens developed just 100,000 yr ago which is a tiny 1/45,000 or 0.002\% of Earth history. 

The number of 'advanced' life forms in the Galaxy is currently impossible to estimate, by lack of statistics.
If the Earth is any guide, the fraction of planets with human-level inhabitants would be the above fraction of 0.002\% of the number of habitable planets of a billion (\S\ref{s:habit}).
The result would be $\sim$22,000, but with a very large uncertainty because the development of humans may well have been a quirk of evolution. 
Natural history on Earth has been strongly influenced by chance events such as meteorite impacts leading to mass extinctions; \citet{kipping2020} has attempted to describe these statistics.
The development of life on Earth soon after its formation suggests that many planets in the habitable zones of their host stars may be inhabited at least by microbes, but any estimate of the number of technologically advanced civilizations capable of interstellar communication or travel is pure speculation at this point.


Even if our Galaxy contains millions of Earth-like planets, observations suggest that very few (if any) of these contain technologically advanced life forms.
Searches for radio signals from extraterrestrial life have been conducted for $>$50 years without any positive result.
In addition, there is no scientific evidence that alien civilizations have ever visited the Earth, which may be a stronger constraint \citep{grimaldi2017}.
The discrepancy between the number of habitable planets and the number of observed civilizations is known as the Fermi paradox, and suggests that evolution of life as it did on Earth is rare (although it is hard to say just how rare).
This section discusses scenarios that aim to explain why this type of evolution may be rare.

One bottleneck for the emergence of life is that some planets may not have the right chemical conditions, even though they lie in the HZ of their host star \citep{krijt2022}. 
Lack of surface water (or other suitable solvent) would be such a case, as the examples of Venus and Mars today indicate.
Isotopic analysis of zircon minerals indicates that water has been present on the Earth's surface for $\sim$4 billion years \citep{genda2016}, which is longer than the oldest fossils, but shorter than the age of the Earth. 
Not much water is needed to start life: the Earth contains $\sim$30 oceans' worth of water, of which 1 on the surface \citep{peslier2017,hirose2021}, which is only 0.023\% by mass, much less than e.g. Europa or Enceladus.
These bodies may actually have too much water: at ocean depths greater than $\sim$100 km, the high pressure turns water into ice, which impedes chemical interaction with the ocean floor \citep{noack2016}.
Another case of chemical inhabitability would be a lack of atmospheric \coo\ or \meth, which provide the greenhouse effect needed around faint young stars.
Lack of nitrogen, phosphorus, or sulphur compounds may also inhibit the build-up of a fully functional biochemistry.

A second bottleneck for the survival and development of life is its ability to adapt to changing conditions on its host planet.
Conditions may change because of seasons and stellar cycles (such as the 11-year sunspot cycle), but also on longer timescales (millions to billions of years) due to orbital changes caused by gravitational interactions with other planets, or long-term stellar evolution. 
For instance, the Sun has brightened by 30\% in the past 4 billion years, i.e. since the start of life on Earth, causing the HZ to move outward by 20\% \citep{mello2023}.
Other causes for changing conditions include continental drift, and interaction with satellites which in our case has led to an increase of the length of day from 20 to 24 hours. 
To deal with such changes, it is crucial that organisms are able to evolve on a timescale matching that of the planetary evolution. 
If the selection pressure is too low, life will see no need to evolve, and will die when conditions change, a condition known as the Gaian bottleneck \citep{chopra2016}

The development of life on other planets may also be limited by extinctions. 
In the fossil record on Earth (which is biased toward animals, a minor fraction of life), at least five mass extinctions are seen since the Cambrian period 0.5 billion years ago.
Most of these are linked to climate changes, following e.g. meteorite impacts or increased volcanic activity.
Pre-Cambrian extinctions are also well documented, for instance following the decrease of \meth\ and rise of \oo\ in our atmosphere \citep{catling2020}.
Since the latter was caused by life itself, this effect is sometimes referred to as the Medean hypothesis.

On Earth, life's key to resilience against extinctions may lie in the development of biodiversity. 
Species A may thrive under current conditions, but species B may do better after conditions change.
Biodiversity naturally develops on planets with a broad range of conditions (temperature, humidity, etc). 
Complex environments will also promote the development of eukaryotic and multi-cellular organisms, which are able to handle many different stimuli at the same time \citep{stevenson2019}.
While multi-cellular species have emerged on several occasions, the evolution of eukaryotes seems to have occurred only once on Earth, suggesting that this was a difficult (hence unlikely) evolutionary invention which may rarely happen on other planets.
The increased information density of complex environments drives the development of systems of interdependent (cooperating) species, which are better able to survive catastrophes than single organisms \citep{adamski2020}. 

\section{Searches for extraterrestrial life}
\label{s:search}

This section discusses ongoing and planned searches for life on other planets.
Several strategies exist for such searches, but most have in common that they seek to identify life by the relatively complex physical and/or chemical structures it produces, which are out of an equilibrium state.
Some searches use the abundances of certain elements or isotopes, others use specific molecules or combinations of molecules.
An example of the latter is the simultaneous presence of \oo\ and \meth, which would under abiotic circumstances react with each other.
Their simultaneous presence in the Earth's atmosphere is caused by their continuous production, respectively by cyanobacteria (as well as plants and trees) and by anaerobic methanogenic microbes (as well as ruminant animals).
Searching for complex organic molecules is not enough: organics with $>$10 atoms have been found in interstellar gas clouds \citep{mcguire2022}, and amino acids have been found in both meteorites (such as Allende) and comets (e.g, 67P) which are clearly lifeless \citep{altwegg2019}.

A different approach involves searches for technologically advanced civilizations, including searches for radio signals with specific patterns such as the Breakthrough Listen project \citep{sheikh2021}, but also for signs of enhanced energy consumption \citep{garrett2015}.
Most searches focus on life as we know it on Earth, because specific predictions can be made what this would look like on other planets.
However, recognizing that extraterrestrial life is probably unlike life on Earth, several groups aim to develop more general biosignatures such as complex patterns in the light reflected from exoplanets \citep{bartlett2022}.
Another idea is to use hub/spoke patterns \citep{walker2020} which are common in biochemical and technological networks \citep{kim2019}, although their observability on exoplanets remains a question.
In any case, the feasibility of `agnostic' biosignatures depend strongly on our ability to directly detect the light reflected or emitted by rocky exoplanets, which may become possible in $\sim$20 years. 

For exoplanets, three types of biosignatures are often considered \citep{schwieterman2018}.
The first type are atmospheric gases such as \oo, the main source of which on Earth is photosynthesis of \hho\ + \coo. 
The \oo\ is a byproduct of the synthesis of organic matter, and may be detected by its A-band, a spectroscopic feature in the far red part of the spectrum. 
This wavelength is accessible to ground-based telescopes, and the detection of \oo\ in exoplanet atmospheres is a science goal of the European Extremely Large Telescope which is starting operations in 2029 in Chile \citep{snellen2013}.
Other searches use \ooo, the photolytic byproduct of \oo, which has a spectral signature at mid-infrared wavelengths, which may be detectable by the future Large Interferometer For Exoplanets \citep{quanz2022}.
The problem with using \oo\ as a biosignature is that it also has an abiotic formation route, namely the photodissociation of \hho\ by ultraviolet starlight, followed by the escape of the light H atoms from the atmosphere \citep{luger2015,schaefer2016}. 
The heavier O atoms stay behind and react with OH into \oo, which therefore cannot be considered an unambiguous biosignature.
Furthermore, this signature would not have worked for the first 2.5 billion years (50\%) of Earth history, when our atmosphere did not contain significant \oo.

The second type of proposed exoplanetary biosignatures are based on light reflected from the surface.
The best known example is the Vegetation Red Edge (VRE), which is an increase in surface reflectance in the red part of the spectrum, due to chlorophyll pigments in plants. 
This feature is commonly used in Earth observation science to measure the level of vegetation in particular areas and seasons.
Like the \oo\ signature above, this signature is strongly biased toward the biochemistry on Earth today.
A recent variation on this theme is the use of circularly polarized light, which is based on the homochirality (single-handedness) of chlorophyll and many other terrestrial biomolecules \citep{patty2021}. 
A less Earth-centric surface biosignature is the use of 'ocean glint', a periodic increase in planetary albedo in the crescent phase of the orbit due to specular reflection off large liquid bodies \citep{lustig2018}.
Strictly speaking, this phenomenon is not a biosignature but a sign of habitable conditions.

A third type of biosignatures for exoplanets are seasonal variations in the abundances of atmospheric gases or surface reflections. 
An example would be the 'Keeling curve', the periodic variation in the Earth's atmospheric \coo\ concentration by 6 ppm (2\%) due to uptake by plants in Spring and Summer, and leaf decay in Fall and Winter.
One caveat of this method is its dependence on the distribution of land mass on a planet. 
The \coo\ periodicity occurs because most land on Earth is on the Northern hemisphere.
For exoplanets, the absence of such a periodicity thus does not mean an absence of plants.

Searches for life within our Solar system usually focus on the planet Mars, the Jovian moon Europa, and Saturn's moons Enceladus and Titan.
The organic molecules found on the Martian surface by the NASA rover Curiosity must have a recent origin, but numerical simulations indicate that delivery by asteroids and comets is a viable source \citep{frantseva2018}.
Curiosity also found seasonal variations in Martian atmospheric \meth\ which were tentatively ascribed to subsurface microbial activity \citep{webster2018}, but more recent measurements did not confirm the seasonality.

The icy moons of Jupiter and Saturn harbour salty oceans below their ice crusts, which are heated by tidal forces. 
Interaction of water with the rocky seafloor could lead to conditions suitable for life, although photosynthesis would not be possible by lack of sunlight.
Recent JWST observations show \coo\ in a geologically young area of Europa, which may originate in the ocean \citep{villanueva2023}.
In the case of Enceladus, water from the subsurface ocean emanates in plumes above the surface.
The NASA/ESA spacecraft Cassini has found salts, \hh, and organics in these plumes, which may be evidence for water-rock interactions on the seafloor under conditions similar to hydrothermal vents on Earth (\citealt{waite2017}; \citealt{postberg2018}). 
The case of Titan is intriguing because this moon harbours the only surface liquid in the Solar system besides Earth; atmospheric pressure is only 50\% higher than on Earth, but temperatures are much lower, and the lakes consist of methane and ethane instead of water.
This moon therefore presents an important example of possibly habitable but very Earth-unlike conditions. 
Missions are underway to Europa (JUICE and Clipper) and planned 2028 to Titan (Dragonfly), and ESA's announcement that its L4 mission will visit Enceladus is great news!


\section{Conclusions and future prospects}
\label{s:concl}

The search for life beyond the Earth is an active field of research, and significant steps have been taken in recent years.
Observations of exoplanets have shown that Earth-like conditions are common in our Galaxy, at least where basic parameters such as mass and size (hence surface gravity) and temperature (based on orbit size and stellar type) are concerned. 
Not all of these planets will have a suitably sized water reservoir, but on the other hand, some exomoons may be habitable, and liquids other than water may be able to support life.
The fraction of rocky planets with atmospheres is unknown, but telescopes to find out are being planned by the American and European space agencies (\citealt{astro2020}; \citealt{voyage2050}), so this number may be known within $\sim$20 years. 
Assuming that life emerges if conditions are suitable, we can say that microbial life therefore is likely to exist outside the Solar system, and cannot be ruled out within.
The development of life soon after the formation of the Earth supports this hypothesis, but we cannot generalize from a single example, which is why the planned space telescopes are important.

Even if the Galaxy may be teeming with life, the number of technologically advanced civilizations appears to be small, based on the lack of radio signals and signs of visits, although firm statistics on this point are lacking.
This discrepancy suggests that the development of life from simple organisms into stable ecosystems is limited by one or more barriers.
The emergence of life on habitable planets may be unlikely, 
the development of sufficient biodiversity may be limited by too low (or too high) selection pressure,
the probability for major evolutionary transitions (such as eukaryotes) may be too low,
or the rate of extinction by catastrophic impacts and other climate changers may be too high.
Besides better insight into the probability of the emergence of life under suitable conditions, what is needed to make progress in this area is a better characterization of rocky exoplanets, especially in terms of their composition and temperature.
Observations of proposed biosignatures on exoplanets will help to test their underlying physics and chemistry, and thus to develop robust unambiguous biosignatures.

The 2020s will see many new facilities, both in space and on the ground, which will make important contributions to this field.
Launched in April 2023, the ESA mission JUICE will arrive in 2031 at the Jovian moon system for exploration of Europa, Ganymede, and Callisto.
In 2024, NASA's Clipper mission has followed, which is targeted toward Europa, and in 2028, Dragonfly will travel to Titan and explore its conditions with drone-type technology.
Scheduled for launch in 2026, the PLATO mission will observe transits of Earth-like planets around Sun-like stars, and allow their basic characterization in terms of temperature and bulk density.
The ARIEL mission will follow in 2028, which will observe transit spectra of mid-size exoplanets, allowing to assess the compositions of their atmospheres.
Meanwhile, the ELT will see its first light in 2029, and started its searches for the \oo\ A-band in giant exoplanet atmospheres and rocky planets around red dwarfs.
Direct imaging of exoplanets will become possible for smaller and older planets with the ELT from the ground and the NASA's Roman telescope (to be launched 2027) from space.
At the end of the decade, ESA's Comet Interceptor (2029) will visit a pristine comet, which has never been close to the Sun before, and inform us about the physical and chemical conditions during the formation of the Solar system.
In the next two decades, NASA's Habitable Worlds Observatory (HWO) will take direct images of rocky exoplanets in the optical, and the Large Interferometer For Exoplanets (LIFE) may take their mid-infrared spectra. 
These latter missions will take significant steps in the characterization of rocky exoplanets, and the testing and development of proposed biosignatures.

While the field of astrobiology enjoys great public interest and significant steps have been made, care should be taken not to make hasty or overstated claims.
Cases such as phosphine on Venus (\citealt{snellen2020}; \citealt{villanueva2021}) or dimethyl sulfide on K2-18b \citep{madhusudhan2023} are recent examples where the public was misled into thinking that alien life had been almost discovered.
By having to retract these claims, we as a field run the risk of losing our credibility and of giving astrobiology a bad name, as happened to the search for exoplanets in the 1970s.
Let us instead focus on solid scientific work and enjoy our wealth of research facilities to study the origin and nature of life in the universe.

\begin{acknowledgement}
The author thanks Mathilde van Dijk for a truly interdisciplinary opportunity, and Sijbren Otto, Tim Lichtenberg, and Wolf Dietrich Geppert for useful comments on a draft version of this chapter.
\end{acknowledgement}

\section*{About the author} 
Floris van der Tak is a Senior Scientist at Space Research Organisation Netherlands (SRON), and Honorary Professor of Astrochemistry and Exoplanet Habitability at the University of Groningen.
He is a Board member of the Netherlands Origins Center and the PRELIFE consortium, and Scientific Coordinator of the EU-funded interdisciplinary research programs oLife and Evolve to investigate the origin of life on Earth and other planets.
As of 2025, he has published 250 articles in peer-reviewed journals, and supervised 77 students and postdocs.
Before his current position, he studied and worked in Bonn (Germany), Leiden (The Netherlands), and Berkeley (United States). 
His research focuses on the formation of stars from interstellar gas clouds, and the habitability of exoplanets.
His students and postdocs use radio and infrared telescopes on Hawaii, in Chile, and in space.

\section*{List of abbreviations}

\begin{table}
\begin{tabular}{ll}
ARIEL & Atmospheric Remote-sensing Infrared Exoplanet Large-survey \\
DNA & Deoxyribo Nucleic Acid \\
ELT & Extremely Large Telescope \\
ESA & European Space Agency \\
HWO & Habitable Worlds Observatory \\
HZ & Habitable Zone \\
JUICE & Jupiter Icy Moons Explorer \\
JWST & James Webb Space Telescope \\
LIFE & Large Interferometer For Exoplanets \\
NASA & National Aeronautics and Space Administration \\
PLATO & Planetary Transits and Oscillations of stars \\
\end{tabular}
\end{table}

\bibliographystyle{spbasic} 
\bibliography{aliens}

\begin{thebibliography}{90}
\providecommand{\natexlab}[1]{#1}
\providecommand{\url}[1]{{#1}}
\providecommand{\urlprefix}{URL }
\expandafter\ifx\csname urlstyle\endcsname\relax
  \providecommand{\doi}[1]{DOI~\discretionary{}{}{}#1}\else
  \providecommand{\doi}{DOI~\discretionary{}{}{}\begingroup
  \urlstyle{rm}\Url}\fi
\providecommand{\eprint}[2][]{\url{#2}}

\bibitem[{{Abbot} et~al.(2012){Abbot}, {Voigt}, {Branson}, {Pierrehumbert},
  {Pollard}, {Le Hir}, and {Koll}}]{abbot2012}
{Abbot} DS, {Voigt} A, {Branson} M, {Pierrehumbert} RT, {Pollard} D, {Le Hir}
  G, {Koll} DDB (2012) {Clouds and Snowball Earth deglaciation}. \grl
  39(20):L20711, \doi{10.1029/2012GL052861}

\bibitem[{Adamski et~al.(2020)Adamski, Eleveld, Sood, Kun, Szil\'agyi,
  Cz\'ar\'an, Szathm\'ary, and Otto}]{adamski2020}
Adamski P, Eleveld M, Sood A, Kun A, Szil\'agyi A, Cz\'ar\'an T, Szathm\'ary E,
  Otto S (2020) From self-replication to replicator systems en route to de novo
  life. Nature Reviews Chemistry 4:386--403, \doi{10.1038/s41570-020-0196-x}

\bibitem[{{Altwegg} et~al.(2019){Altwegg}, {Balsiger}, and
  {Fuselier}}]{altwegg2019}
{Altwegg} K, {Balsiger} H, {Fuselier} SA (2019) {Cometary Chemistry and the
  Origin of Icy Solar System Bodies: The View After Rosetta}. \araa
  57:113--155, \doi{10.1146/annurev-astro-091918-104409}, \eprint{1908.04046}

\bibitem[{{Anglada-Escud{\'e}} et~al.(2016){Anglada-Escud{\'e}}, {Amado},
  {Barnes}, {Berdi{\~n}as}, {Butler}, {Coleman}, {de La Cueva}, {Dreizler},
  {Endl}, {Giesers}, {Jeffers}, {Jenkins}, {Jones}, {Kiraga}, {K{\"u}rster},
  {L{\'o}pez-Gonz{\'a}lez}, {Marvin}, {Morales}, {Morin}, {Nelson}, {Ortiz},
  {Ofir}, {Paardekooper}, {Reiners}, {Rodr{\'\i}guez},
  {Rodr{\'\i}guez-L{\'o}pez}, {Sarmiento}, {Strachan}, {Tsapras}, {Tuomi}, and
  {Zechmeister}}]{anglada2016}
{Anglada-Escud{\'e}} G, {Amado} PJ, {Barnes} J, {Berdi{\~n}as} ZM, {Butler} RP,
  {Coleman} GAL, {de La Cueva} I, {Dreizler} S, {Endl} M, {Giesers} B,
  {Jeffers} SV, {Jenkins} JS, {Jones} HRA, {Kiraga} M, {K{\"u}rster} M,
  {L{\'o}pez-Gonz{\'a}lez} MJ, {Marvin} CJ, {Morales} N, {Morin} J, {Nelson}
  RP, {Ortiz} JL, {Ofir} A, {Paardekooper} SJ, {Reiners} A, {Rodr{\'\i}guez} E,
  {Rodr{\'\i}guez-L{\'o}pez} C, {Sarmiento} LF, {Strachan} JP, {Tsapras} Y,
  {Tuomi} M, {Zechmeister} M (2016) {A terrestrial planet candidate in a
  temperate orbit around Proxima Centauri}. \nat 536(7617):437--440,
  \doi{10.1038/nature19106}, \eprint{1609.03449}

\bibitem[{{Bains} et~al.(2021){Bains}, {Petkowski}, {Zhan}, and
  {Seager}}]{bains2021}
{Bains} W, {Petkowski} JJ, {Zhan} Z, {Seager} S (2021) {Evaluating Alternatives
  to Water as Solvents for Life: The Example of Sulfuric Acid}. Life 11(5):400,
  \doi{10.3390/life11050400}

\bibitem[{{Bartlett} et~al.(2022){Bartlett}, {Li}, {Gu}, {Sinapayen}, {Fan},
  {Natraj}, {Jiang}, {Crisp}, and {Yung}}]{bartlett2022}
{Bartlett} S, {Li} J, {Gu} L, {Sinapayen} L, {Fan} S, {Natraj} V, {Jiang} JH,
  {Crisp} D, {Yung} YL (2022) {Assessing planetary complexity and potential
  agnostic biosignatures using epsilon machines}. Nature Astronomy 6:387--392,
  \doi{10.1038/s41550-021-01559-x}, \eprint{2202.03699}

\bibitem[{{Bell} et~al.(2023){Bell}, {Welbanks}, {Schlawin}, {Line}, {Fortney},
  {Greene}, {Ohno}, {Parmentier}, {Rauscher}, {Beatty}, {Mukherjee}, {Wiser},
  {Boyer}, {Rieke}, and {Stansberry}}]{bell2023}
{Bell} TJ, {Welbanks} L, {Schlawin} E, {Line} MR, {Fortney} JJ, {Greene} TP,
  {Ohno} K, {Parmentier} V, {Rauscher} E, {Beatty} TG, {Mukherjee} S, {Wiser}
  LS, {Boyer} ML, {Rieke} MJ, {Stansberry} JA (2023) {Methane throughout the
  atmosphere of the warm exoplanet WASP-80b}. \nat 623(7988):709--712,
  \doi{10.1038/s41586-023-06687-0}, \eprint{2309.04042}

\bibitem[{{Broadley} et~al.(2022){Broadley}, {Bekaert}, {Piani}, {F{\"u}ri},
  and {Marty}}]{broadley2022}
{Broadley} MW, {Bekaert} DV, {Piani} L, {F{\"u}ri} E, {Marty} B (2022) {Origin
  of life-forming volatile elements in the inner Solar System}. Nature
  611(7935):245--255, \doi{10.1038/s41586-022-05276-x}

\bibitem[{Bruno(1584)}]{bruno1584}
Bruno G (1584) De l'infinito universo et mondi

\bibitem[{{Buchhave} et~al.(2012){Buchhave}, {Latham}, {Johansen}, {Bizzarro},
  {Torres}, {Rowe}, {Batalha}, {Borucki}, {Brugamyer}, {Caldwell}, {Bryson},
  {Ciardi}, {Cochran}, {Endl}, {Esquerdo}, {Ford}, {Geary}, {Gilliland},
  {Hansen}, {Isaacson}, {Laird}, {Lucas}, {Marcy}, {Morse}, {Robertson},
  {Shporer}, {Stefanik}, {Still}, and {Quinn}}]{buchhave2012}
{Buchhave} LA, {Latham} DW, {Johansen} A, {Bizzarro} M, {Torres} G, {Rowe} JF,
  {Batalha} NM, {Borucki} WJ, {Brugamyer} E, {Caldwell} C, {Bryson} ST,
  {Ciardi} DR, {Cochran} WD, {Endl} M, {Esquerdo} GA, {Ford} EB, {Geary} JC,
  {Gilliland} RL, {Hansen} T, {Isaacson} H, {Laird} JB, {Lucas} PW, {Marcy} GW,
  {Morse} JA, {Robertson} P, {Shporer} A, {Stefanik} RP, {Still} M, {Quinn} SN
  (2012) {An abundance of small exoplanets around stars with a wide range of
  metallicities}. \nat 486(7403):375--377, \doi{10.1038/nature11121}

\bibitem[{{Catling} and {Zahnle}(2020)}]{catling2020}
{Catling} DC, {Zahnle} KJ (2020) {The Archean atmosphere}. Science Advances
  6(9):eaax1420, \doi{10.1126/sciadv.aax1420}

\bibitem[{{Charbonneau} et~al.(2000){Charbonneau}, {Brown}, {Latham}, and
  {Mayor}}]{charbonneau2000}
{Charbonneau} D, {Brown} TM, {Latham} DW, {Mayor} M (2000) {Detection of
  Planetary Transits Across a Sun-like Star}. \apjl 529(1):L45--L48,
  \doi{10.1086/312457}, \eprint{astro-ph/9911436}

\bibitem[{{Chopra} and {Lineweaver}(2016)}]{chopra2016}
{Chopra} A, {Lineweaver} CH (2016) {The Case for a Gaian Bottleneck: The
  Biology of Habitability}. Astrobiology 16(1):7--22,
  \doi{10.1089/ast.2015.1387}

\bibitem[{Cleaves et~al.(2023)Cleaves, Hystad, Prabhu, Wong, Cody, Economon,
  and Hazen}]{cleaves2023}
Cleaves HJ, Hystad G, Prabhu A, Wong ML, Cody GD, Economon S, Hazen RM (2023) A
  robust, agnostic molecular biosignature based on machine learning.
  Proceedings of the National Academy of Sciences 120(41):e2307149120,
  \doi{10.1073/pnas.2307149120}

\bibitem[{{Cockell} et~al.(2016){Cockell}, {Bush}, {Bryce}, {Direito},
  {Fox-Powell}, {Harrison}, {Lammer}, {Landenmark}, {Martin-Torres},
  {Nicholson}, {Noack}, {O'Malley-James}, {Payler}, {Rushby}, {Samuels},
  {Schwendner}, {Wadsworth}, and {Zorzano}}]{cockell2016}
{Cockell} CS, {Bush} T, {Bryce} C, {Direito} S, {Fox-Powell} M, {Harrison} JP,
  {Lammer} H, {Landenmark} H, {Martin-Torres} J, {Nicholson} N, {Noack} L,
  {O'Malley-James} J, {Payler} SJ, {Rushby} A, {Samuels} T, {Schwendner} P,
  {Wadsworth} J, {Zorzano} MP (2016) {Habitability: A Review}. Astrobiology
  16(1):89--117, \doi{10.1089/ast.2015.1295}

\bibitem[{{Crossfield}(2015)}]{crossfield2015}
{Crossfield} IJM (2015) {Observations of Exoplanet Atmospheres}. \pasp
  127(956):941, \doi{10.1086/683115}, \eprint{1507.03966}

\bibitem[{{Dayal} et~al.(2015){Dayal}, {Cockell}, {Rice}, and
  {Mazumdar}}]{dayal2015}
{Dayal} P, {Cockell} C, {Rice} K, {Mazumdar} A (2015) {The Quest for Cradles of
  Life: Using the Fundamental Metallicity Relation to Hunt for the Most
  Habitable Type of Galaxy}. \apjl 810(1):L2, \doi{10.1088/2041-8205/810/1/L2},
  \eprint{1507.04346}

\bibitem[{{Deming} et~al.(2019){Deming}, {Louie}, and {Sheets}}]{deming2019}
{Deming} D, {Louie} D, {Sheets} H (2019) {How to Characterize the Atmosphere of
  a Transiting Exoplanet}. \pasp 131(995):013001,
  \doi{10.1088/1538-3873/aae5c5}, \eprint{1810.04175}

\bibitem[{{Dobos} et~al.(2022){Dobos}, {Haris}, {Kamp}, and {van der
  Tak}}]{dobos2022}
{Dobos} V, {Haris} A, {Kamp} IEE, {van der Tak} FFS (2022) {A target list for
  searching for habitable exomoons}. \mnras 513(4):5290--5298,
  \doi{10.1093/mnras/stac1180}, \eprint{2204.11614}

\bibitem[{{Ducrot} et~al.(2025){Ducrot}, {Lagage}, {Min}, {Gillon}, {Bell},
  {Tremblin}, {Greene}, {Dyrek}, {Bouwman}, {Waters}, {G{\"u}del}, {Henning},
  {Vandenbussche}, {Absil}, {Barrado}, {Boccaletti}, {Coulais}, {Decin},
  {Edwards}, {Gastaud}, {Glasse}, {Kendrew}, {Olofsson}, {Patapis}, {Pye},
  {Rouan}, {Whiteford}, {Argyriou}, {Cossou}, {Glauser}, {Krause}, {Lahuis},
  {Royer}, {Scheithauer}, {Colina}, {van Dishoeck}, {Ostlin}, {Ray}, and
  {Wright}}]{ducrot2025}
{Ducrot} E, {Lagage} PO, {Min} M, {Gillon} M, {Bell} TJ, {Tremblin} P, {Greene}
  T, {Dyrek} A, {Bouwman} J, {Waters} R, {G{\"u}del} M, {Henning} T,
  {Vandenbussche} B, {Absil} O, {Barrado} D, {Boccaletti} A, {Coulais} A,
  {Decin} L, {Edwards} B, {Gastaud} R, {Glasse} A, {Kendrew} S, {Olofsson} G,
  {Patapis} P, {Pye} J, {Rouan} D, {Whiteford} N, {Argyriou} I, {Cossou} C,
  {Glauser} AM, {Krause} O, {Lahuis} F, {Royer} P, {Scheithauer} S, {Colina} L,
  {van Dishoeck} EF, {Ostlin} G, {Ray} TP, {Wright} G (2025) {Combined analysis
  of the 12.8 and 15 {\ensuremath{\mu}}m JWST/MIRI eclipse observations of
  TRAPPIST-1 b}. Nature Astronomy 9:358--369, \doi{10.1038/s41550-024-02428-z},
  \eprint{2412.11627}

\bibitem[{ESA(2021)}]{voyage2050}
ESA (2021) Voyage 2050 report. \url{https://www.cosmos.esa.int/web/voyage-2050}

\bibitem[{{Ferlet} et~al.(1987){Ferlet}, {Hobbs}, and
  {Vidal-Madjar}}]{ferlet1987}
{Ferlet} R, {Hobbs} LM, {Vidal-Madjar} A (1987) {The beta Pictoris
  circumstellar disk. V. Time variations of the Ca II-K line.} \aap
  185:267--270

\bibitem[{{Frantseva} et~al.(2018){Frantseva}, {Mueller}, {ten Kate}, {van der
  Tak}, and {Greenstreet}}]{frantseva2018}
{Frantseva} K, {Mueller} M, {ten Kate} IL, {van der Tak} FFS, {Greenstreet} S
  (2018) {Delivery of organics to Mars through asteroid and comet impacts}.
  \icarus 309:125--133, \doi{10.1016/j.icarus.2018.03.006}, \eprint{1803.03270}

\bibitem[{{Gaia Collaboration}(2023)}]{gaia2023}
{Gaia Collaboration} (2023) {Gaia Data Release 3. Stellar multiplicity, a
  teaser for the hidden treasure}. \aap 674:A34,
  \doi{10.1051/0004-6361/202243782}, \eprint{2206.05595}

\bibitem[{{Garrett}(2015)}]{garrett2015}
{Garrett} MA (2015) {Application of the mid-IR radio correlation to the {\^{G}}
  sample and the search for advanced extraterrestrial civilisations}. \aap
  581:L5, \doi{10.1051/0004-6361/201526687}, \eprint{1508.02624}

\bibitem[{{Genda}(2016)}]{genda2016}
{Genda} H (2016) {Origin of Earth's oceans: An assessment of the total amount,
  history and supply of water}. Geochemical Journal 50(1):27--42,
  \doi{10.2343/geochemj.2.0398}

\bibitem[{{Gillon} et~al.(2017){Gillon}, {Triaud}, {Demory}, {Jehin}, {Agol},
  {Deck}, {Lederer}, {de Wit}, {Burdanov}, {Ingalls}, {Bolmont}, {Leconte},
  {Raymond}, {Selsis}, {Turbet}, {Barkaoui}, {Burgasser}, {Burleigh}, {Carey},
  {Chaushev}, {Copperwheat}, {Delrez}, {Fernandes}, {Holdsworth}, {Kotze}, {Van
  Grootel}, {Almleaky}, {Benkhaldoun}, {Magain}, and {Queloz}}]{gillon2017}
{Gillon} M, {Triaud} AHMJ, {Demory} BO, {Jehin} E, {Agol} E, {Deck} KM,
  {Lederer} SM, {de Wit} J, {Burdanov} A, {Ingalls} JG, {Bolmont} E, {Leconte}
  J, {Raymond} SN, {Selsis} F, {Turbet} M, {Barkaoui} K, {Burgasser} A,
  {Burleigh} MR, {Carey} SJ, {Chaushev} A, {Copperwheat} CM, {Delrez} L,
  {Fernandes} CS, {Holdsworth} DL, {Kotze} EJ, {Van Grootel} V, {Almleaky} Y,
  {Benkhaldoun} Z, {Magain} P, {Queloz} D (2017) {Seven temperate terrestrial
  planets around the nearby ultracool dwarf star TRAPPIST-1}. \nat
  542(7642):456--460, \doi{10.1038/nature21360}, \eprint{1703.01424}

\bibitem[{{Greaves} et~al.(1998){Greaves}, {Holland}, {Moriarty-Schieven},
  {Jenness}, {Dent}, {Zuckerman}, {McCarthy}, {Webb}, {Butner}, {Gear}, and
  {Walker}}]{greaves1998}
{Greaves} JS, {Holland} WS, {Moriarty-Schieven} G, {Jenness} T, {Dent} WRF,
  {Zuckerman} B, {McCarthy} C, {Webb} RA, {Butner} HM, {Gear} WK, {Walker} HJ
  (1998) {A Dust Ring around ? Eridani: Analog to the Young Solar System}.
  \apjl 506(2):L133--L137, \doi{10.1086/311652}, \eprint{astro-ph/9808224}

\bibitem[{{Grimaldi}(2017)}]{grimaldi2017}
{Grimaldi} C (2017) {Signal coverage approach to the detection probability of
  hypothetical extraterrestrial emitters in the Milky Way}. Scientific Reports
  7:46273, \doi{10.1038/srep46273}, \eprint{1704.04028}

\bibitem[{Herschel(1795)}]{herschel1795}
Herschel F (1795) Philosophical Transactions of the Royal Society

\bibitem[{{Hirose} et~al.(2021){Hirose}, {Wood}, and
  {Vo\v{c}adlo}}]{hirose2021}
{Hirose} K, {Wood} B, {Vo\v{c}adlo} L (2021) Light elements in the earth's
  core. Nature Reviews Earth \& Environment 2(9):645--658,
  \doi{10.1038/s43017-021-00203-6}

\bibitem[{{Hu} et~al.(2024){Hu}, {Bello-Arufe}, {Zhang}, {Paragas},
  {Zilinskas}, {van Buchem}, {Bess}, {Patel}, {Ito}, {Damiano}, {Scheucher},
  {Oza}, {Knutson}, {Miguel}, {Dragomir}, {Brandeker}, and {Demory}}]{hu2024}
{Hu} R, {Bello-Arufe} A, {Zhang} M, {Paragas} K, {Zilinskas} M, {van Buchem} C,
  {Bess} M, {Patel} J, {Ito} Y, {Damiano} M, {Scheucher} M, {Oza} AV, {Knutson}
  HA, {Miguel} Y, {Dragomir} D, {Brandeker} A, {Demory} BO (2024) {A secondary
  atmosphere on the rocky exoplanet 55 Cancri e}. \nat 630(8017):609--612,
  \doi{10.1038/s41586-024-07432-x}, \eprint{2405.04744}

\bibitem[{Huygens(1698)}]{huygens1698}
Huygens C (1698) Kosmotheoros, sive De terris coelestibus earumque ornatu
  conjecturae

\bibitem[{{Javaux}(2019)}]{javaux2019}
{Javaux} EJ (2019) {Challenges in evidencing the earliest traces of life}. \nat
  572(7770):451--460, \doi{10.1038/s41586-019-1436-4}

\bibitem[{{Kalas} et~al.(2005){Kalas}, {Graham}, and {Clampin}}]{kalas2005}
{Kalas} P, {Graham} JR, {Clampin} M (2005) {A planetary system as the origin of
  structure in Fomalhaut's dust belt}. \nat 435(7045):1067--1070,
  \doi{10.1038/nature03601}, \eprint{astro-ph/0506574}

\bibitem[{Kant(1755)}]{kant1755}
Kant I (1755) Allgemeine Naturgeschichte und Theorie des Himmels

\bibitem[{{Kasting} et~al.(1993){Kasting}, {Whitmire}, and
  {Reynolds}}]{kasting1993}
{Kasting} JF, {Whitmire} DP, {Reynolds} RT (1993) {Habitable Zones around Main
  Sequence Stars}. \icarus 101(1):108--128, \doi{10.1006/icar.1993.1010}

\bibitem[{{Kenworthy} et~al.(2023){Kenworthy}, {Lock}, {Kennedy}, {van
  Capelleveen}, {Mamajek}, {Carone}, {Hambsch}, {Masiero}, {Mainzer},
  {Kirkpatrick}, {Gomez}, {Leinhardt}, {Dou}, {Tanna}, {Sainio}, {Barker},
  {Charbonnel}, {Garde}, {Le D{\^u}}, {Mulato}, {Petit}, and {Rizzo
  Smith}}]{kenworthy2023}
{Kenworthy} M, {Lock} S, {Kennedy} G, {van Capelleveen} R, {Mamajek} E,
  {Carone} L, {Hambsch} FJ, {Masiero} J, {Mainzer} A, {Kirkpatrick} JD, {Gomez}
  E, {Leinhardt} Z, {Dou} J, {Tanna} P, {Sainio} A, {Barker} H, {Charbonnel} S,
  {Garde} O, {Le D{\^u}} P, {Mulato} L, {Petit} T, {Rizzo Smith} M (2023) {A
  planetary collision afterglow and transit of the resultant debris cloud}.
  \nat 622(7982):251--254, \doi{10.1038/s41586-023-06573-9},
  \eprint{2310.08360}

\bibitem[{{Kenworthy} and {Mamajek}(2015)}]{kenworthy2015}
{Kenworthy} MA, {Mamajek} EE (2015) {Modeling Giant Extrasolar Ring Systems in
  Eclipse and the Case of J1407b: Sculpting by Exomoons?} \apj 800(2):126,
  \doi{10.1088/0004-637X/800/2/126}, \eprint{1501.05652}

\bibitem[{Kim et~al.(2019)Kim, Smith, Mathis, Raymond, and Walker}]{kim2019}
Kim H, Smith HB, Mathis C, Raymond J, Walker SI (2019) Universal scaling across
  biochemical networks on earth. Science Advances 5(1):eaau0149,
  \doi{10.1126/sciadv.aau0149},
  \urlprefix\url{https://www.science.org/doi/abs/10.1126/sciadv.aau0149},
  \eprint{https://www.science.org/doi/pdf/10.1126/sciadv.aau0149}

\bibitem[{{Kipping}(2020)}]{kipping2020}
{Kipping} D (2020) {An objective Bayesian analysis of life's early start and
  our late arrival}. Proceedings of the National Academy of Science
  117(22):11995--12003, \doi{10.1073/pnas.1921655117}, \eprint{2005.09008}

\bibitem[{{Kipping} et~al.(2022){Kipping}, {Bryson}, {Burke}, {Christiansen},
  {Hardegree-Ullman}, {Quarles}, {Hansen}, {Szul{\'a}gyi}, and
  {Teachey}}]{kipping2022}
{Kipping} D, {Bryson} S, {Burke} C, {Christiansen} J, {Hardegree-Ullman} K,
  {Quarles} B, {Hansen} B, {Szul{\'a}gyi} J, {Teachey} A (2022) {An exomoon
  survey of 70 cool giant exoplanets and the new candidate Kepler-1708 b-i}.
  Nature Astronomy 6:367--380, \doi{10.1038/s41550-021-01539-1},
  \eprint{2201.04643}

\bibitem[{{Kite} and {Barnett}(2020)}]{kite2020}
{Kite} ES, {Barnett} MN (2020) {Exoplanet secondary atmosphere loss and
  revival}. Proceedings of the National Academy of Science 117:18264--18271,
  \doi{10.1073/pnas.2006177117}, \eprint{2006.02589}

\bibitem[{{Kopparapu} et~al.(2013){Kopparapu}, {Ramirez}, {Kasting}, {Eymet},
  {Robinson}, {Mahadevan}, {Terrien}, {Domagal-Goldman}, {Meadows}, and
  {Deshpande}}]{kopparapu2013}
{Kopparapu} RK, {Ramirez} R, {Kasting} JF, {Eymet} V, {Robinson} TD,
  {Mahadevan} S, {Terrien} RC, {Domagal-Goldman} S, {Meadows} V, {Deshpande} R
  (2013) {Habitable Zones around Main-sequence Stars: New Estimates}. \apj
  765(2):131, \doi{10.1088/0004-637X/765/2/131}, \eprint{1301.6674}

\bibitem[{{Krijt} et~al.(2022){Krijt}, {Kama}, {McClure}, {Teske}, {Bergin},
  {Shorttle}, {Walsh}, and {Raymond}}]{krijt2022}
{Krijt} S, {Kama} M, {McClure} M, {Teske} J, {Bergin} EA, {Shorttle} O, {Walsh}
  KJ, {Raymond} SN (2022) {Chemical Habitability: Supply and Retention of
  Life's Essential Elements During Planet Formation}. arXiv e-prints
  arXiv:2203.10056, \doi{10.48550/arXiv.2203.10056}, \eprint{2203.10056}

\bibitem[{{Lineweaver} et~al.(2004){Lineweaver}, {Fenner}, and
  {Gibson}}]{lineweaver2004}
{Lineweaver} CH, {Fenner} Y, {Gibson} BK (2004) {The Galactic Habitable Zone
  and the Age Distribution of Complex Life in the Milky Way}. Science
  303(5654):59--62, \doi{10.1126/science.1092322}, \eprint{astro-ph/0401024}

\bibitem[{{Lodders}(2003)}]{lodders2003}
{Lodders} K (2003) {Solar System Abundances and Condensation Temperatures of
  the Elements}. \apj 591(2):1220--1247, \doi{10.1086/375492}

\bibitem[{{Luger} and {Barnes}(2015)}]{luger2015}
{Luger} R, {Barnes} R (2015) {Extreme Water Loss and Abiotic O2Buildup on
  Planets Throughout the Habitable Zones of M Dwarfs}. Astrobiology
  15(2):119--143, \doi{10.1089/ast.2014.1231}, \eprint{1411.7412}

\bibitem[{{Lustig-Yaeger} et~al.(2018){Lustig-Yaeger}, {Meadows}, {Tovar
  Mendoza}, {Schwieterman}, {Fujii}, {Luger}, and {Robinson}}]{lustig2018}
{Lustig-Yaeger} J, {Meadows} VS, {Tovar Mendoza} G, {Schwieterman} EW, {Fujii}
  Y, {Luger} R, {Robinson} TD (2018) {Detecting Ocean Glint on Exoplanets Using
  Multiphase Mapping}. \aj 156(6):301, \doi{10.3847/1538-3881/aaed3a},
  \eprint{1901.05011}

\bibitem[{{Madhusudhan}(2019)}]{madhusudhan2019}
{Madhusudhan} N (2019) {Exoplanetary Atmospheres: Key Insights, Challenges, and
  Prospects}. \araa 57:617--663, \doi{10.1146/annurev-astro-081817-051846},
  \eprint{1904.03190}

\bibitem[{{Madhusudhan} et~al.(2023){Madhusudhan}, {Sarkar}, {Constantinou},
  {Holmberg}, {Piette}, and {Moses}}]{madhusudhan2023}
{Madhusudhan} N, {Sarkar} S, {Constantinou} S, {Holmberg} M, {Piette} AAA,
  {Moses} JI (2023) {Carbon-bearing Molecules in a Possible Hycean Atmosphere}.
  \apjl 956(1):L13, \doi{10.3847/2041-8213/acf577}, \eprint{2309.05566}

\bibitem[{{Marois} et~al.(2008){Marois}, {Macintosh}, {Barman}, {Zuckerman},
  {Song}, {Patience}, {Lafreni{\`e}re}, and {Doyon}}]{marois2008}
{Marois} C, {Macintosh} B, {Barman} T, {Zuckerman} B, {Song} I, {Patience} J,
  {Lafreni{\`e}re} D, {Doyon} R (2008) {Direct Imaging of Multiple Planets
  Orbiting the Star HR 8799}. Science 322(5906):1348,
  \doi{10.1126/science.1166585}, \eprint{0811.2606}

\bibitem[{{Mastrogiuseppe} et~al.(2019){Mastrogiuseppe}, {Poggiali}, {Hayes},
  {Lunine}, {Seu}, {Mitri}, and {Lorenz}}]{mastrogiuseppe2019}
{Mastrogiuseppe} M, {Poggiali} V, {Hayes} AG, {Lunine} JI, {Seu} R, {Mitri} G,
  {Lorenz} RD (2019) {Deep and methane-rich lakes on Titan}. Nature Astronomy
  3:535--542, \doi{10.1038/s41550-019-0714-2}

\bibitem[{{Mayor} and {Queloz}(1995)}]{mayor1995}
{Mayor} M, {Queloz} D (1995) {A Jupiter-mass companion to a solar-type star}.
  Nature 378(6555):355--359, \doi{10.1038/378355a0}

\bibitem[{{McDonough} and {Arevalo}(2008)}]{mcdonough2008}
{McDonough} WF, {Arevalo} J Ricardo (2008) {Uncertainties in the composition of
  Earth, its core and silicate sphere}. In: Journal of Physics Conference
  Series, Journal of Physics Conference Series, vol 136, p 022006,
  \doi{10.1088/1742-6596/136/2/022006}

\bibitem[{{McGuire}(2022)}]{mcguire2022}
{McGuire} BA (2022) {2021 Census of Interstellar, Circumstellar, Extragalactic,
  Protoplanetary Disk, and Exoplanetary Molecules}. \apjs 259(2):30,
  \doi{10.3847/1538-4365/ac2a48}, \eprint{2109.13848}

\bibitem[{{Meadows} and {Barnes}(2018)}]{meadows2018}
{Meadows} VS, {Barnes} RK (2018) {Factors Affecting Exoplanet Habitability}.
  In: {Deeg} HJ, {Belmonte} JA (eds) Handbook of Exoplanets, p~57,
  \doi{10.1007/978-3-319-55333-757}

\bibitem[{{Mello} and {Fria{\c{c}}a}(2023)}]{mello2023}
{Mello} FdS, {Fria{\c{c}}a} ACS (2023) {Planetary geodynamics and age
  constraints on circumstellar habitable zones around main sequence stars}.
  International Journal of Astrobiology 22(4):272--316,
  \doi{10.1017/S1473550423000083}

\bibitem[{Merino et~al.(2019)Merino, Aronson, Bojanova, Feyhl-Buska, Wong,
  Zhang, and Giovannelli}]{merino2019}
Merino N, Aronson HS, Bojanova DP, Feyhl-Buska J, Wong ML, Zhang S, Giovannelli
  D (2019) Living at the extremes: Extremophiles and the limits of life in a
  planetary context. Frontiers in Microbiology 10,
  \doi{10.3389/fmicb.2019.00780},
  \urlprefix\url{https://www.frontiersin.org/articles/10.3389/fmicb.2019.00780}

\bibitem[{{Millero} et~al.(2008){Millero}, {Feistel}, {Wright}, and
  {McDougall}}]{millero2008}
{Millero} FJ, {Feistel} R, {Wright} DG, {McDougall} TJ (2008) {The composition
  of Standard Seawater and the definition of the Reference-Composition Salinity
  Scale}. Deep Sea Research Part I: Oceanographic Research 55(1):50--72,
  \doi{10.1016/j.dsr.2007.10.001}

\bibitem[{{Miret-Roig} et~al.(2022){Miret-Roig}, {Bouy}, {Raymond}, {Tamura},
  {Bertin}, {Barrado}, {Olivares}, {Galli}, {Cuillandre}, {Sarro}, {Berihuete},
  and {Hu{\'e}lamo}}]{miretroig2022}
{Miret-Roig} N, {Bouy} H, {Raymond} SN, {Tamura} M, {Bertin} E, {Barrado} D,
  {Olivares} J, {Galli} PAB, {Cuillandre} JC, {Sarro} LM, {Berihuete} A,
  {Hu{\'e}lamo} N (2022) {A rich population of free-floating planets in the
  Upper Scorpius young stellar association}. Nature Astronomy 6:89--97,
  \doi{10.1038/s41550-021-01513-x}, \eprint{2112.11999}

\bibitem[{NAS(2022)}]{astro2020}
NAS (2022) Pathways to discovery in astronomy and astrophysics for the 2020s.
  \url{https://science.nasa.gov/astrophysics/resources/decadal-survey/2020-decadal-survey/}

\bibitem[{NASA(2025)}]{xpl-arxiv}
NASA (2025) Exoplanet archive. \url{https://exoplanetarchive.ipac.caltech.edu/}

\bibitem[{Newton(1713)}]{newton1713}
Newton I (1713) Scholium Generale, an Appendix to the 2nd edition of
  Philosophiae Naturalis Principia Mathematica

\bibitem[{{Noack} et~al.(2016){Noack}, {H{\"o}ning}, {Rivoldini},
  {Heistracher}, {Zimov}, {Journaux}, {Lammer}, {Van Hoolst}, and
  {Bredeh{\"o}ft}}]{noack2016}
{Noack} L, {H{\"o}ning} D, {Rivoldini} A, {Heistracher} C, {Zimov} N,
  {Journaux} B, {Lammer} H, {Van Hoolst} T, {Bredeh{\"o}ft} JH (2016)
  {Water-rich planets: How habitable is a water layer deeper than on Earth?}
  \icarus 277:215--236, \doi{10.1016/j.icarus.2016.05.009}

\bibitem[{{Oosterloo} et~al.(2021){Oosterloo}, {H{\"o}ning}, {Kamp}, and {van
  der Tak}}]{oosterloo2021}
{Oosterloo} M, {H{\"o}ning} D, {Kamp} IEE, {van der Tak} FFS (2021) {The role
  of planetary interior in the long-term evolution of atmospheric CO$_{2}$ on
  Earth-like exoplanets}. \aap 649:A15, \doi{10.1051/0004-6361/202039664},
  \eprint{2103.09505}

\bibitem[{{Patty} et~al.(2021){Patty}, {K{\"u}hn}, {Lambrev}, {Spadaccia},
  {Jens Hoeijmakers}, {Keller}, {Mulder}, {Pallichadath}, {Poch}, {Snik},
  {Stam}, {Pommerol}, and {Demory}}]{patty2021}
{Patty} CHL, {K{\"u}hn} JG, {Lambrev} PH, {Spadaccia} S, {Jens Hoeijmakers} H,
  {Keller} C, {Mulder} W, {Pallichadath} V, {Poch} O, {Snik} F, {Stam} DM,
  {Pommerol} A, {Demory} BO (2021) {Biosignatures of the Earth. I. Airborne
  spectropolarimetric detection of photosynthetic life}. \aap 651:A68,
  \doi{10.1051/0004-6361/202140845}, \eprint{2106.00493}

\bibitem[{{Perryman}(2018)}]{perryman2018}
{Perryman} M (2018) {The Exoplanet Handbook}. Cambridge University Press

\bibitem[{{Peslier} et~al.(2017){Peslier}, {Sch{\"o}nb{\"a}chler}, {Busemann},
  and {Karato}}]{peslier2017}
{Peslier} AH, {Sch{\"o}nb{\"a}chler} M, {Busemann} H, {Karato} SI (2017) {Water
  in the Earth's Interior: Distribution and Origin}. \ssr 212(1-2):743--810,
  \doi{10.1007/s11214-017-0387-z}

\bibitem[{{Petkowski} et~al.(2020){Petkowski}, {Bains}, and
  {Seager}}]{petkowski2020}
{Petkowski} JJ, {Bains} W, {Seager} S (2020) {On the Potential of Silicon as a
  Building Block for Life}. Life 10(6):84, \doi{10.3390/life10060084}

\bibitem[{{Postberg} et~al.(2018){Postberg}, {Khawaja}, {Abel}, {Choblet},
  {Glein}, {Gudipati}, {Henderson}, {Hsu}, {Kempf}, {Klenner},
  {Moragas-Klostermeyer}, {Magee}, {N{\"o}lle}, {Perry}, {Reviol}, {Schmidt},
  {Srama}, {Stolz}, {Tobie}, {Trieloff}, and {Waite}}]{postberg2018}
{Postberg} F, {Khawaja} N, {Abel} B, {Choblet} G, {Glein} CR, {Gudipati} MS,
  {Henderson} BL, {Hsu} HW, {Kempf} S, {Klenner} F, {Moragas-Klostermeyer} G,
  {Magee} B, {N{\"o}lle} L, {Perry} M, {Reviol} R, {Schmidt} J, {Srama} R,
  {Stolz} F, {Tobie} G, {Trieloff} M, {Waite} JH (2018) {Macromolecular organic
  compound s from the depths of Enceladus}. Nature 558(7711):564--568,
  \doi{10.1038/s41586-018-0246-4}

\bibitem[{{Quanz} et~al.(2022){Quanz}, {Ottiger}, {Fontanet}, {Kammerer},
  {Menti}, {Dannert}, {Gheorghe}, {Absil}, {Airapetian}, {Alei}, {Allart},
  {Angerhausen}, {Blumenthal}, {Buchhave}, {Cabrera},
  {Carri{\'o}n-Gonz{\'a}lez}, {Chauvin}, {Danchi}, {Dandumont}, {Defr{\'e}re},
  {Dorn}, {Ehrenreich}, {Ertel}, {Fridlund}, {Garc{\'\i}a Mu{\~n}oz},
  {Gasc{\'o}n}, {Girard}, {Glauser}, {Grenfell}, {Guidi}, {Hagelberg},
  {Helled}, {Ireland}, {Janson}, {Kopparapu}, {Korth}, {Kozakis}, {Kraus},
  {L{\'e}ger}, {Leedj{\"a}rv}, {Lichtenberg}, {Lillo-Box}, {Linz}, {Liseau},
  {Loicq}, {Mahendra}, {Malbet}, {Mathew}, {Mennesson}, {Meyer}, {Mishra},
  {Molaverdikhani}, {Noack}, {Oza}, {Pall{\'e}}, {Parviainen}, {Quirrenbach},
  {Rauer}, {Ribas}, {Rice}, {Romagnolo}, {Rugheimer}, {Schwieterman},
  {Serabyn}, {Sharma}, {Stassun}, {Szul{\'a}gyi}, {Wang}, {Wunderlich},
  {Wyatt}, and {LIFE Collaboration}}]{quanz2022}
{Quanz} SP, {Ottiger} M, {Fontanet} E, {Kammerer} J, {Menti} F, {Dannert} F,
  {Gheorghe} A, {Absil} O, {Airapetian} VS, {Alei} E, {Allart} R, {Angerhausen}
  D, {Blumenthal} S, {Buchhave} LA, {Cabrera} J, {Carri{\'o}n-Gonz{\'a}lez}
  {\'O}, {Chauvin} G, {Danchi} WC, {Dandumont} C, {Defr{\'e}re} D, {Dorn} C,
  {Ehrenreich} D, {Ertel} S, {Fridlund} M, {Garc{\'\i}a Mu{\~n}oz} A,
  {Gasc{\'o}n} C, {Girard} JH, {Glauser} A, {Grenfell} JL, {Guidi} G,
  {Hagelberg} J, {Helled} R, {Ireland} MJ, {Janson} M, {Kopparapu} RK, {Korth}
  J, {Kozakis} T, {Kraus} S, {L{\'e}ger} A, {Leedj{\"a}rv} L, {Lichtenberg} T,
  {Lillo-Box} J, {Linz} H, {Liseau} R, {Loicq} J, {Mahendra} V, {Malbet} F,
  {Mathew} J, {Mennesson} B, {Meyer} MR, {Mishra} L, {Molaverdikhani} K,
  {Noack} L, {Oza} AV, {Pall{\'e}} E, {Parviainen} H, {Quirrenbach} A, {Rauer}
  H, {Ribas} I, {Rice} M, {Romagnolo} A, {Rugheimer} S, {Schwieterman} EW,
  {Serabyn} E, {Sharma} S, {Stassun} KG, {Szul{\'a}gyi} J, {Wang} HS,
  {Wunderlich} F, {Wyatt} MC, {LIFE Collaboration} (2022) {Large Interferometer
  For Exoplanets (LIFE). I. Improved exoplanet detection yield estimates for a
  large mid-infrared space-interferometer mission}. \aap 664:A21,
  \doi{10.1051/0004-6361/202140366}, \eprint{2101.07500}

\bibitem[{{Rudnick} and {Gao}(2003)}]{rudnick2003}
{Rudnick} RL, {Gao} S (2003) {Composition of the Continental Crust}. Treatise
  on Geochemistry 3:659, \doi{10.1016/B0-08-043751-6/03016-4}

\bibitem[{{Schaefer} et~al.(2016){Schaefer}, {Wordsworth}, {Berta-Thompson},
  and {Sasselov}}]{schaefer2016}
{Schaefer} L, {Wordsworth} RD, {Berta-Thompson} Z, {Sasselov} D (2016)
  {Predictions of the Atmospheric Composition of GJ 1132b}. \apj 829(2):63,
  \doi{10.3847/0004-637X/829/2/63}, \eprint{1607.03906}

\bibitem[{{Schwieterman} et~al.(2018){Schwieterman}, {Kiang}, {Parenteau},
  {Harman}, {DasSarma}, {Fisher}, {Arney}, {Hartnett}, {Reinhard}, {Olson},
  {Meadows}, {Cockell}, {Walker}, {Grenfell}, {Hegde}, {Rugheimer}, {Hu}, and
  {Lyons}}]{schwieterman2018}
{Schwieterman} EW, {Kiang} NY, {Parenteau} MN, {Harman} CE, {DasSarma} S,
  {Fisher} TM, {Arney} GN, {Hartnett} HE, {Reinhard} CT, {Olson} SL, {Meadows}
  VS, {Cockell} CS, {Walker} SI, {Grenfell} JL, {Hegde} S, {Rugheimer} S, {Hu}
  R, {Lyons} TW (2018) {Exoplanet Biosignatures: A Review of Remotely
  Detectable Signs of Life}. Astrobiology 18(6):663--708,
  \doi{10.1089/ast.2017.1729}, \eprint{1705.05791}

\bibitem[{{Sheikh} et~al.(2021){Sheikh}, {Smith}, {Price}, {DeBoer}, {Lacki},
  {Czech}, {Croft}, {Gajjar}, {Isaacson}, {Lebofsky}, {MacMahon}, {Ng},
  {Perez}, {Siemion}, {Webb}, {Zic}, {Drew}, and {Worden}}]{sheikh2021}
{Sheikh} SZ, {Smith} S, {Price} DC, {DeBoer} D, {Lacki} BC, {Czech} DJ, {Croft}
  S, {Gajjar} V, {Isaacson} H, {Lebofsky} M, {MacMahon} DHE, {Ng} C, {Perez}
  KI, {Siemion} APV, {Webb} CI, {Zic} A, {Drew} J, {Worden} SP (2021) {Analysis
  of the Breakthrough Listen signal of interest blc1 with a technosignature
  verification framework}. Nature Astronomy 5:1153--1162,
  \doi{10.1038/s41550-021-01508-8}, \eprint{2111.06350}

\bibitem[{{Snellen} et~al.(2013){Snellen}, {de Kok}, {le Poole}, {Brogi}, and
  {Birkby}}]{snellen2013}
{Snellen} IAG, {de Kok} RJ, {le Poole} R, {Brogi} M, {Birkby} J (2013) {Finding
  Extraterrestrial Life Using Ground-based High-dispersion Spectroscopy}. \apj
  764(2):182, \doi{10.1088/0004-637X/764/2/182}, \eprint{1302.3251}

\bibitem[{{Snellen} et~al.(2020){Snellen}, {Guzman-Ramirez}, {Hogerheijde},
  {Hygate}, and {van der Tak}}]{snellen2020}
{Snellen} IAG, {Guzman-Ramirez} L, {Hogerheijde} MR, {Hygate} APS, {van der
  Tak} FFS (2020) {Re-analysis of the 267 GHz ALMA observations of Venus. No
  statistically significant detection of phosphine}. \aap 644:L2,
  \doi{10.1051/0004-6361/202039717}, \eprint{2010.09761}

\bibitem[{{Stevenson} and {Large}(2019)}]{stevenson2019}
{Stevenson} DS, {Large} S (2019) {Evolutionary exobiology: towards the
  qualitative assessment of biological potential on exoplanets}. International
  Journal of Astrobiology 18(3):204--208, \doi{10.1017/S1473550417000349}

\bibitem[{{Str{\o}m} et~al.(2020){Str{\o}m}, {Bodewits}, {Knight}, {Kiefer},
  {Jones}, {Kral}, {Matr{\`a}}, {Bodman}, {Capria}, {Cleeves}, {Fitzsimmons},
  {Haghighipour}, {Harrison}, {Iglesias}, {Kama}, {Linnartz}, {Majumdar}, {de
  Mooij}, {Milam}, {Opitom}, {Rebollido}, {Rogers}, {Snodgrass}, {Sousa-Silva},
  {Xu}, {Lin}, and {Zieba}}]{strom2020}
{Str{\o}m} PA, {Bodewits} D, {Knight} MM, {Kiefer} F, {Jones} GH, {Kral} Q,
  {Matr{\`a}} L, {Bodman} E, {Capria} MT, {Cleeves} I, {Fitzsimmons} A,
  {Haghighipour} N, {Harrison} JHD, {Iglesias} D, {Kama} M, {Linnartz} H,
  {Majumdar} L, {de Mooij} EJW, {Milam} SN, {Opitom} C, {Rebollido} I, {Rogers}
  LK, {Snodgrass} C, {Sousa-Silva} C, {Xu} S, {Lin} ZY, {Zieba} S (2020)
  {Exocomets from a Solar System Perspective}. \pasp 132(1016):101001,
  \doi{10.1088/1538-3873/aba6a0}, \eprint{2007.09155}

\bibitem[{{Suer} et~al.(2023){Suer}, {Jackson}, {Grewal}, {Dalou}, and
  {Lichtenberg}}]{suer2023}
{Suer} TA, {Jackson} C, {Grewal} DS, {Dalou} C, {Lichtenberg} T (2023) {The
  distribution of volatile elements during rocky planet formation}. Frontiers
  in Earth Science 11:1159412, \doi{10.3389/feart.2023.1159412},
  \eprint{2311.18262}

\bibitem[{Swedenborg(1734)}]{swedenborg1734}
Swedenborg E (1734) Principia rerum naturalium

\bibitem[{{Tsai} et~al.(2023){Tsai}, {Lee}, and {Powell}}]{tsai2023}
{Tsai} SM, {Lee} EKH, {Powell} ea Diana (2023) {Photochemically produced
  SO$_{2}$ in the atmosphere of WASP-39b}. Nature 617(7961):483--487,
  \doi{10.1038/s41586-023-05902-2}, \eprint{2211.10490}

\bibitem[{{Villanueva} et~al.(2021){Villanueva}, {Cordiner}, {Irwin}, {de
  Pater}, {Butler}, {Gurwell}, {Milam}, {Nixon}, {Luszcz-Cook}, {Wilson},
  {Kofman}, {Liuzzi}, {Faggi}, {Fauchez}, {Lippi}, {Cosentino}, {Thelen},
  {Moullet}, {Hartogh}, {Molter}, {Charnley}, {Arney}, {Mandell}, {Biver},
  {Vandaele}, {de Kleer}, and {Kopparapu}}]{villanueva2021}
{Villanueva} GL, {Cordiner} M, {Irwin} PGJ, {de Pater} I, {Butler} B, {Gurwell}
  M, {Milam} SN, {Nixon} CA, {Luszcz-Cook} SH, {Wilson} CF, {Kofman} V,
  {Liuzzi} G, {Faggi} S, {Fauchez} TJ, {Lippi} M, {Cosentino} R, {Thelen} AE,
  {Moullet} A, {Hartogh} P, {Molter} EM, {Charnley} S, {Arney} GN, {Mandell}
  AM, {Biver} N, {Vandaele} AC, {de Kleer} KR, {Kopparapu} R (2021) {No
  evidence of phosphine in the atmosphere of Venus from independent analyses}.
  Nature Astronomy 5:631--635, \doi{10.1038/s41550-021-01422-z}

\bibitem[{{Villanueva} et~al.(2023){Villanueva}, {Hammel}, {Milam}, {Faggi},
  {Kofman}, {Roth}, {Hand}, {Paganini}, {Stansberry}, {Spencer}, {Protopapa},
  {Strazzulla}, {Cruz-Mermy}, {Glein}, {Cartwright}, and
  {Liuzzi}}]{villanueva2023}
{Villanueva} GL, {Hammel} HB, {Milam} SN, {Faggi} S, {Kofman} V, {Roth} L,
  {Hand} KP, {Paganini} L, {Stansberry} J, {Spencer} J, {Protopapa} S,
  {Strazzulla} G, {Cruz-Mermy} G, {Glein} CR, {Cartwright} R, {Liuzzi} G (2023)
  {Endogenous CO$_{2}$ ice mixture on the surface of Europa and no detection of
  plume activity}. Science 381(6664):1305--1308, \doi{10.1126/science.adg4270}

\bibitem[{{Wade} et~al.(2021){Wade}, {Byrne}, {Ballentine}, and
  {Drakesmith}}]{wade2021}
{Wade} J, {Byrne} DJ, {Ballentine} CJ, {Drakesmith} H (2021) {Temporal
  variation of planetary iron as a driver of evolution}. Proceedings of the
  National Academy of Science 118(51):e2109865118,
  \doi{10.1073/pnas.2109865118}

\bibitem[{{Waite} et~al.(2017){Waite}, {Glein}, {Perryman}, {Teolis}, {Magee},
  {Miller}, {Grimes}, {Perry}, {Miller}, {Bouquet}, {Lunine}, {Brockwell}, and
  {Bolton}}]{waite2017}
{Waite} JH, {Glein} CR, {Perryman} RS, {Teolis} BD, {Magee} BA, {Miller} G,
  {Grimes} J, {Perry} ME, {Miller} KE, {Bouquet} A, {Lunine} JI, {Brockwell} T,
  {Bolton} SJ (2017) {Cassini finds molecular hydrogen in the Enceladus plume:
  Evidence for hydrothermal processes}. Science 356(6334):155--159,
  \doi{10.1126/science.aai8703}

\bibitem[{{Walker} et~al.(2020){Walker}, {Cronin}, {Drew}, {Domagal-Goldman},
  {Fisher}, {Line}, and {Millsaps}}]{walker2020}
{Walker} SI, {Cronin} L, {Drew} A, {Domagal-Goldman} S, {Fisher} T, {Line} M,
  {Millsaps} C (2020) {Probabilistic Frameworks for Life Detection}. In:
  {Meadows} VS, {Arney} GN, {Schmidt} BE, {Des Marais} DJ (eds) Planetary
  Astrobiology, p 477, \doi{10.2458/azu-uapress-9780816540068}

\bibitem[{{Webster} et~al.(2018){Webster}, {Mahaffy}, and
  {Atreya}}]{webster2018}
{Webster} CR, {Mahaffy} PR, {Atreya} ea Sushil~K (2018) {Background levels of
  methane in Mars{\textquoteright} atmosphere show strong seasonal variations}.
  Science 360(6393):1093--1096, \doi{10.1126/science.aaq0131}

\bibitem[{{Zieba} et~al.(2019){Zieba}, {Zwintz}, {Kenworthy}, and
  {Kennedy}}]{zwieba2019}
{Zieba} S, {Zwintz} K, {Kenworthy} MA, {Kennedy} GM (2019) {Transiting
  exocomets detected in broadband light by TESS in the {\ensuremath{\beta}}
  Pictoris system}. \aap 625:L13, \doi{10.1051/0004-6361/201935552},
  \eprint{1903.11071}

\end{thebibliography}
\end{document}